\documentclass[prl,superscriptaddress,twocolumn,showpacs]{revtex4}

\usepackage{graphicx}% Include figure files
\usepackage{amsmath} 
\usepackage{amssymb}

\begin{document}

\title{Finite-time effects and ultraweak ergodicity breaking in
superdiffusive dynamics}

\author{Alja\v z Godec}
\email{aljaz.godec@ki.si}
\affiliation{Institute for Physics \& Astronomy, University of Potsdam,
14476 Potsdam-Golm, Germany}
\affiliation{National Institute of Chemistry, 1000 Ljubljana, Slovenia}
\author{Ralf Metzler}
\email{rmetzler@uni-potsdam.de}
\affiliation{Institute for Physics \& Astronomy, University of Potsdam,
14476 Potsdam-Golm, Germany}
\affiliation{Physics Department, Tampere University of Technology, 33101,
Tampere, Finland}

\date{\today}

\begin{abstract}
We  study the ergodic properties of superdiffusive, spatiotemporally
coupled L{\'e}vy walk processes. For trajectories of finite duration,
we reveal a distinct scatter of the scaling exponents of the time
averaged mean squared displacement $\overline{\delta^2}$ around the
ensemble value $3-\alpha$ ($1<\alpha<2$) ranging from ballistic motion
to subdiffusion, in strong contrast to the behavior of subdiffusive
processes. In addition we find a significant dependence of the
trajectory-to-trajectory average of $\overline{\delta^2}$ as function
of the finite measurement time. This so-called finite-time amplitude
depression and the scatter of the scaling exponent is vital in the
quantitative evaluation of superdiffusive processes. Comparing the
long time average of the  second moment with the ensemble mean squared
displacement, these only differ  by a  constant factor,
an ultraweak ergodicity breaking.
\end{abstract}


\maketitle

Suppose you are recording the trajectories of individual
blue sharks in the ocean over time $t$. Calculating the time
averaged mean squared displacement (MSD) $\overline{\delta^2}$ for each shark
you find that some animals move almost ballistically while others appear
to move much slower. Does this indicate that the animals follow different
generic motion patterns? As we show here for the celebrated
L{\'e}vy walk (LW) model of superdiffusion, the intrinsic non-ergodicity
in trajectories of finite length indeed gives rise to a wide distribution of
apparent scaling exponents, even to subdiffusive values, although the motion is
produced from identical
distributions. This surprising finding is accompanied by a strong reduction
of the amplitude of $\overline{\delta^2}$ at finite measurement times and
strongly contrasts the non-ergodicity observed in subdiffusive motion.

Blue sharks are indeed just one example of marine predators followed over
large distances that show scaling laws in their foraging behavior consistent
with LW dynamics \cite{sims}, similar to findings from other tracking studies
of individual animals or humans \cite{nathan,bara,spider}. LWs are more widely
applied, \emph{inter alia\/} to describe intermittent chaotic systems
\cite{Zumofen1,Zumofen2,geisel0}, turbulent flow \cite{swinney}, accelerated
diffusion in Josephson junctions \cite{Geisel}, negative Hall-resistance in
semiconductors \cite{Geisel2}, diffusion of atoms in optical lattices
\cite{Marksteiner} and of light in disordered media \cite{wiersma}, blinking
statistics of quantum dots \cite{Silbey}, movement strategies in mussels
\cite{deJager}, or even T-cell motility in the brain \cite{Harris}. Many
of these systems may be analyzed on the single trajectory level.

Despite this ubiquity of LWs their ergodic behavior has not been studied in
detail. However the question whether a system is ergodic becomes relevant
when instead of the conventional MSD $\langle x^2(t)\rangle=\int x^2P(x,t)dx$
defined as ensemble average over the probability density $P(x,t)$ we use time
averages over single trajectories. For time series $x(t)$ of duration $T$ the
time averaged MSD is defined via
\begin{equation}
\label{tamsd}
\overline{\delta x^2}=\frac{1}{T-\tau}\int_{0}^{T-\tau}
\Big[x(t+\tau)-x(t)\Big]^2dt,
\end{equation}
where $\tau$ denotes the lag time. The behavior of $\overline{\delta x^2}$
has been studied in detail for the subdiffusion case, $\langle
x^2(t)\rangle \simeq
t^{\gamma}$ with $0<\gamma<1$, revealing distinct discrepancies between
ensemble and time averaged MSD for scale free waiting time  processes
\cite{pt,yonghe}. This so-called weak ergodicity breaking (WEB) means that
$\langle x^2(\tau)\rangle\neq\overline{\delta x^2}$ even for long $T$
\cite{pt,yonghe}, while other subdiffusive processes such as fractional
Brownian motion are ergodic in the sense that $\langle x^2(\tau)\rangle=
\overline{\delta x^2}$ for sufficiently long $T$ \cite{deng,jae}. WEB has
indeed been observed in experiments, for instance, for the motion of
protein channels in the walls of living cells \cite{weigel} and of lipid
granules in yeast cells \cite{lene}.

To study the ergodic properties of of LWs we recall their definition
within continuous time random walk (CTRW) theory \cite{shleklawong}.
A CTRW is based on the joint distribution $\Psi(x,t)$. For each jump
we draw from $\Psi(x,t)$ a random waiting time $t$ and jump length
$x$ \cite{montroll,klablushle,Klafter2}. To describe superdiffusive
processes $\langle x^2(t)\rangle\simeq t^{\gamma}$ with $\gamma>1$, LWs
are endowed with a spatiotemporal coupling for which we chose the simplest
form $\Psi(x,t)=\frac{1}{2}\psi(t)\delta(|x|-vt)$ \cite{klablushle}. Confined
by an expanding horizon at positions $\pm vt$ from the origin, this CTRW
performs statistically independent free paths with constant velocity $|v|$,
whose durations are distributed according to the power law $\psi(t)=\int
\Psi(x,t)dx\sim t^{-(1+\alpha)}$. For $0<\alpha<1$ the resulting motion is
ballistic, $\gamma=2$, for $1<\alpha<2$ we observe sub-ballistic
superdiffusion with $\gamma=3-\alpha$, while for $\alpha>2$ the motion
is normal diffusive, $\gamma=1$ \cite{Zumofen1,Zumofen2}. The mean
sojourn time $\langle t\rangle=\int_0^{\infty}t\psi(t)dt$ is infinite
for $0<\alpha<1$ and finite otherwise. In contrast to L{\'e}vy flights
with their diverging variance \cite{report}, LWs are thus physical models for
particles with a maximum propagation speed. Apart from the description in terms
of above continuous time random walk  scheme with $\Psi(x,t)$, LWs can be
described as  a renewal  process \cite{Geisel}, in  terms of  a master
equation \cite{Trefan}, a fractional transport equation \cite{Ralf15},
or a Langevin approach based on subordination \cite{Marcin}.

Here we focus on the  behavior of time averages and ergodic properties
of  LWs  in  the  relevant  superdiffusive  range  $1<\alpha<2$.  From
analytical  results and extensive  numerical simulations  we highlight
the particular role of the finiteness of trajectories when calculating
the  time averages. Namely,  we show  that the scaling exponents of
$\overline{\delta^2}$ apparently become random quantities and that the
amplitude of  the time averages is a function of the  measurement time
$T$.  Moreover,  we report  an  \emph{ultraweak\/} ergodicity breaking
of superdiffusive  LWs. These effects are important to interpret
time averages of LW processes.

A full analytical solution for the the time averaged MSD $\overline{\delta
x^2}$ is obtained from the renewal framework \cite{Geisel}. The starting
point is the velocity autocorrelation function $C_{v}(t)=\lim_{t'\to\infty}
|t'-t|^{-1}\int_0^{|t'-t|}v(t'')v(t''+t)dt''$, where the time average is taken
over a trajectory of infinite length.  In the  velocity model for LWs
employed here the velocity fluctuates between $+v$ and $-v$ with equal
probability,   meaning   that  only   single   events  contribute   to
$C_{v}(t)$,  which in turn  is the  result of  an averaging  of event
durations along a trajectory. The problem can be rephrased in terms of
the probability that  a walker is in an ongoing  event of duration $T$
between $0$ and $t$ given that we pick an arbitrary origin on the time
axis.  To obtain  $C_{v}$ we  simply average  over all  such possible
durations. Once $C_{v}$ is  known, $\overline{\delta x^2}$ is readily
obtained    from     the    Green-Kubo    formula    \cite{Green_Kubo}
$\overline{\delta x^2(\tau)} =  2\int_0^{\tau} (\tau-t) C_ {v}(t)dt$.
For infinite trajectories, we obtain the result
\begin{equation}
\overline{\delta x^2(\tau)}= 2\left(\frac{(1+\tau)^{3-\alpha}-1}{(3-\alpha)
(2-\alpha)}-\frac{\tau}{2-\alpha}\right), 
\label{eq1}
\end{equation}
where we have  set $|v|=1$. As for $1<\alpha<2$  the mean waiting time
$\langle t\rangle$ is  finite, individual trajectories at sufficiently
long (infinite) times become  self-averaging, such that there will be
no  difference  between  $\overline{\delta x^2(\tau)}$  obtained  from
different   trajectories  and the   trajectory-to-trajectory  averaged
quantity $\langle\overline{\delta  x^2(\tau)}\rangle$. In other words,
the  actual series of  events is  irrelevant in  the case  of infinite
trajectories.  In reality one  never deals with infinite trajectories,
albeit they might become extremely  long. Once a trajectory is finite,
irrespective  of its  actual  length, there  always  exits a  non-zero
probability that the walker will  be `locked' in a single motion event
persisting  along   a  great  fraction  or  even   during  the  entire
trajectory. Not  surprisingly, the MSDs  $\overline{\delta x^2(\tau)}$
of individual  trajectories will  not coincide but  show a  scatter of
amplitudes, as shown below.

\begin{figure}
\includegraphics[width=0.48\textwidth]{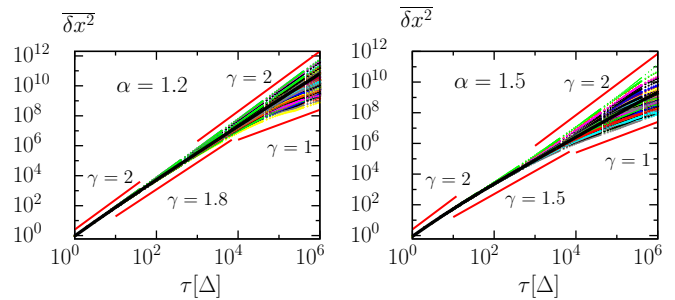}
\caption{Time averaged mean squared displacement $\overline{\delta x^2}$
obtained from 400 trajectories for $\alpha=1.2$ (top) and $\alpha= 1.5$
(bottom). Full red lines depict different scaling behaviors as
indicated, the full black lines are averages over all trajectories.}
\label{fg1}
\end{figure} 

In our simulations we use the concrete form $\psi(t)=\alpha(1+t)^{-(1+
\alpha)}$ for the waiting time distribution. From this asymptotic
power-law we generate $M=10^4$ time series of particle coordinates
$x_j(t)$, where $j$ labels different trajectories. We calculate the
ensemble averaged MSD $\langle\Delta x^2(\tau)\rangle=M^{-1}\sum_{j=1}^M
x_j(\tau)^2$ and the time averaged MSD through Eq.~(\ref{tamsd}).
Fig.~\ref{fg1} shows typical results for $\overline{\delta x^2}$ for 400
different trajectories of duration $T=10^8$ time steps ($\Delta$) for
$\alpha=1.2$
and $\alpha=1.5$. Remarkably, while $\overline{\delta x^2}$ for all
trajectories coincides and shows superdiffusive scaling at shorter lag times,
at longer $\tau$, $\overline{\delta x^2}$ displays a wide spread of slopes
ranging from ballistic motion to subdiffusion ($\gamma<1$). At the same
time the ensemble-averaged MSD predicts a unique long-time scaling of
the form $\langle x^2(t)\rangle\simeq t^{3- \alpha}$, confirmed by our
simulations (not shown here). Thus ergodicity, the equivalence of long
time and ensemble average is broken. Moreover, self-averaging does not
take place. In contrast to subdiffusive CTRW with diverging mean waiting
time, where the scaling is identical for all trajectories but the generalized
diffusion coefficient becomes a random variable \cite{yonghe,pt}, here we
observe that the scaling exponent of individual trajectories appears
random. We note that this effect is not due to bad
statistics at larger $\tau$ as $\tau\ll T$ is fulfilled for all $\tau$
shown in Fig.~\ref{fg1}. Performing an average over all trajectories,
$\langle\overline{\delta x^2}\rangle$, the full black lines in Fig.~\ref{fg1},
the result seems to follow the scaling predicted by Eq.~(\ref{eq1}).
However, this agreement is only apparent, see below.

\begin{figure}
\includegraphics[width=0.3\textwidth]{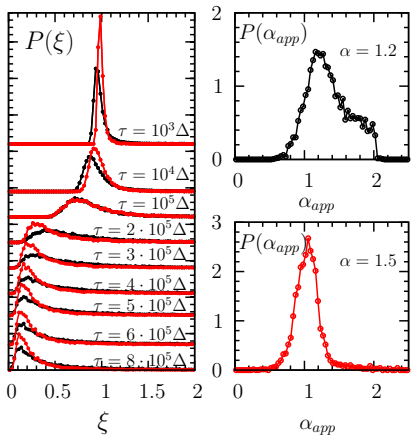}
\caption{Left: Scatter distribution $P(\xi)$ of the time averaged MSD
$\overline{\delta x^2}$ versus $\xi=\overline{\delta x^2}/\langle\overline{
\delta x^2}\rangle$ for $\alpha=1.2$ (black) and $\alpha=1.5$ (red). Right:
Scatter distribution $P(\alpha_{\mathrm{app}})$ of the apparent scaling
exponent $\alpha_{\mathrm{app}}$ obtained by least-squares fit of
$\overline{\delta x^2}\sim t^{\alpha_{\mathrm{app}}}$ to the last decade of
the time series in Fig.~\ref{fg1}.}
\label{fg2}
\end{figure}

We first quantify the deviations between different trajectories in
terms of the distribution of $\overline{\delta^2}$ around the
trajectory-to-trajectory average $\langle\overline{\delta x^2}\rangle$,
\begin{equation}
P(\xi|\tau)=\left<\delta\left(\frac{\overline{\delta x^2(\tau)}}{\langle
\overline{\delta x^2(\tau)}\rangle}-\xi\right)\right>.
\label{eq2}
\end{equation}
The results for $\alpha=1.2$ and $\alpha=1.5$ are shown in Fig.~\ref{fg2}
(Left). In the case of an infinite trajectory we would find a sharp peak
at $\xi=1$, which is approximately observed for the shortest $\tau$. In
contrast for finite-time trajectories $P(\xi|\tau)$ apparently relaxes
towards a skewed limiting distribution with a maximum well below the
ergodic value $\xi=1$. Therefore the average value appears to be dominated
by one or few very long waiting time events locked onto a given velocity
mode, and the self-averaging is not fulfilled. In addition we measure
the long-time scaling in individual trajectories by least-squares fit of
the last decade of $\overline{\delta  x^2}$ to the power-law $t^{\alpha_{
\mathrm{app}}}$, obtaining the scatter distribution of the \emph{apparent\/}
scaling exponent $\alpha_\mathrm{app}$. The resulting distributions $P(\alpha_{
\mathrm{app}})$  for  $\alpha=1.2$ and $\alpha=1.5$ are shown in
Fig.~\ref{fg2} (Right). We see that the maximum of $P(\alpha_{\mathrm{app}})$
is well below the infinite-time average $\alpha_\mathrm{app}=\alpha$, and
that a non-negligible fraction of trajectories in fact exhibits subdiffusion.
This demonstrates that the time average of a superdiffusive dynamical process
can in fact display subdiffusive behavior on the level of single trajectories
of finite duration. Again, we see that finite time averages such as $\overline{
\delta^2}$ are obviously dominated by either extremely long motion events
pushing $\alpha_\mathrm{app}$ to values closer to $\alpha_\mathrm{app}=2$,
while strong oscillations between velocity modes induce localization effects
and values $\alpha_\mathrm{app}<1$. These observations will be crucial for
the correct interpretation of single trajectory measurements of superdiffusive
processes.

\begin{figure}
\includegraphics[width=0.48\textwidth]{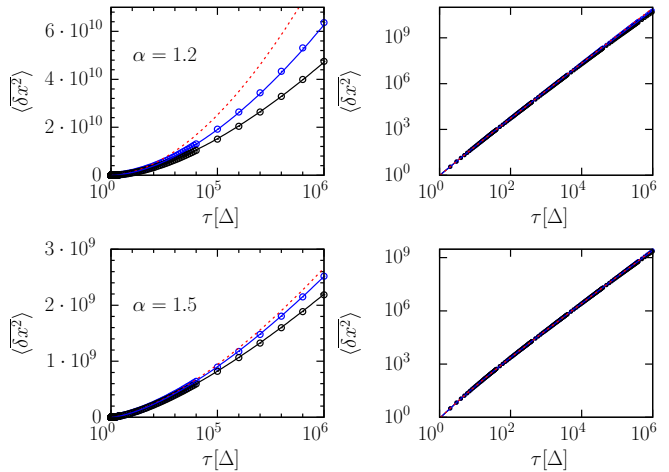}
\caption{$\langle\overline{\delta  x^2}\rangle$ (dashed red line) and
$\langle\overline{\delta x^2}\rangle_f$ (full lines) with $T=10^7\Delta$
(black symbols) and $T=10^8\Delta$ (blue symbols). The right panel shows
the same plots on the logarithmic scale.}
\label{fg3}
\end{figure}

\begin{figure}
\includegraphics[width=0.3\textwidth]{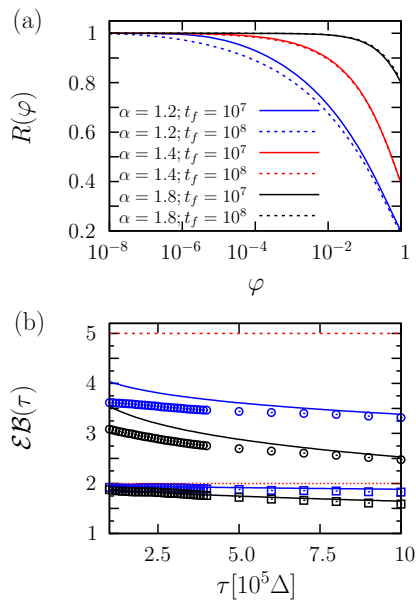}
\caption{(a) $R(\varphi)$ for various values of $\alpha$ and $T$;
(b) Ergodicity breaking parameter as function of lag time, $\mathcal{EB}
(\tau)$ for $\alpha=1.2$ (circles) and $\alpha=1.5$ (squares) for $T=10^7
\Delta$ (black) and $T=10^8\Delta$ (blue). The dashed and dotted red lines
denote the theoretical values for infinite-time trajectories, and the full
lines correspond to the predictions for the finite-time case.}
\label{fg4}
\end{figure}

Having established that there is no unique scaling of $\overline{\delta
x^2}$ along finite-time single trajectories one might wonder whether and
how the finiteness of single trajectories affects the corresponding
average over an ensemble of trajectories. This problem can be treated
exactly with the renewal approach. Once an arbitrary origin is specified
on the time axis the probability that the walker is in a motion event of
duration $\vartheta$ at time 0 is $p_0(\vartheta)d\vartheta=\vartheta\psi(
\vartheta)d\vartheta/\overline{\vartheta}$, where $\overline{\vartheta}$ is
the average time span of $\vartheta$ along a finite-time trajectory of
duration $T$ and ensures the correct normalization, $\overline{\vartheta}=
\int_0^{T}\vartheta\psi(\vartheta)d\vartheta\equiv\frac{1}{\alpha-1}
(1-\alpha(1+\alpha  T)(1+T)^{-\alpha})$. The probability that the event
persists until $t$ is $p_p(t|\vartheta)=(\vartheta-t)/\vartheta$, such that
the probability that the walker is in a motion event of duration $\vartheta$
between 0 and t is $p_0(\vartheta)p_p(t|\vartheta)d\vartheta$. The velocity
autocorrelation function $C_{v}^f(t)$ is then obtained by averaging over
all possible durations up to $T$, such that
\begin{equation}
C_{v}^f(t)=\frac{(1+t)^{1-\alpha}+(1+T)^{1-\alpha}((\alpha-1)t-(1+\alpha
T))}{1-\alpha(1+\alpha T)(1+T)^{-\alpha}}
\label{eq3}
\end{equation} 
for $t<T$, such that for finite-time trajectories we find
\begin{equation}
\left<\overline{\delta x^2}\right>_f=2\frac{\frac{(1+\tau)^{3-\alpha}
-1}{(3-\alpha)(2-\alpha)}-\frac{t}{2-\alpha}+\left(\frac{(\alpha-1)
\tau^3}{6(1+T)^{\alpha}}-\frac{(1+\alpha T)\tau^2}{2(1+T)^{\alpha}}
\right)}{(\alpha-1)\overline{\vartheta}}.
\label{eq4}
\end{equation}
For long $T$ the time averaged MSD has the form
\begin{equation}
\left<\overline{\delta x^2}\right>_f\sim\left<\overline{\delta x^2}(\tau)
\right>+T^{3-\alpha}\left[\frac{\alpha-1}{3}\left(\frac{t}{T}\right)^3-
\alpha\left(\frac{t}{T}\right)^2\right].
\label{eq5}
\end{equation}
Simulations results for $\langle\overline{\delta x^2}\rangle$ are shown
in Fig.~\ref{fg3}, demonstrating good agreement with the result (\ref{eq4}).
Indeed
we find that on a logarithmic scale (Right, the conventional representation
of time averaged MSD data) one hardly observes deviations from Eq.~(\ref{eq1}),
however, on a linear scale pronounced deviations are apparent (Left). Of course
as $t/T\to1$ these deviations would become increasingly pronounced also on
the logarithmic scale. To asses the importance
of correction terms for given values of $\alpha$ and $T$ it is instructive to
consider the ratio of $\langle\overline{\delta^2}\rangle>$ for finite-time
and infinite-time trajectories,
\begin{equation}
R=\left<\overline{\delta x^2}\right>_f\Big/\left<\overline{\delta x^2}\right>
\end{equation}
as function of the relative time-lag $\varphi\equiv\tau/T$. The results for
various cases are shown in Fig.~\ref{fg4} (Left). Thus, when $\alpha=1.2$ for
instance, $R(\varphi)$ decreases to 0.2 as $\varphi$ approaches 1. This
\emph{finite-time depression\/} is important in relating the amplitude of
the measured time averaged MSD to the anomalous diffusion coefficient of
the process. In the Brownian limit $\alpha=2$, $R(\varphi)$ is independent
of the finite measurement time $T$ and equals 1.

Finally we investigate the nature of the ergodicity breaking and, in
particular, the role of the finiteness of trajectories. As already
noted by Zumofen and Klafter \cite{Zumofen2} the time averaged MSD
differs from the corresponding ensemble average, thus ergodicity is
broken. We define the ergodicity-breaking parameter $\mathcal{EB}(\tau)=
\langle\overline{\delta x^2}\rangle/\langle\Delta x^2(\tau)\rangle$ as
ratio of time versus ensemble averaged MSD \cite{REM}. For our choice of
$\psi(x,t)$
the ensemble averaged MSD asymptotically is $\langle\Delta x^2(\tau)\rangle
\sim2(\alpha-1)\tau^{3-\alpha}/((3-\alpha)(2-\alpha))$ for $1<\alpha<2$ as
$\tau\to\infty$ \cite{Zumofen2}. Thus $\mathcal{EB}(\tau)=1/(\alpha-1)$ as
$\tau\to\infty$, that is, time and ensemble averages differ only in terms
of a constant. We call this phenomenon\textit{ultraweak ergodicity breaking},
in contrast to the stronger weak ergodicity breaking of scale-free subdiffusive
processes. According  to
Eq.~(\ref{eq4}) we expect that the finiteness of trajectories will also
affect  $\mathcal{EB}(\tau)$. Interestingly $\mathcal{EB}(\tau)$
appears  to  be almost  independent  of $\tau$,  as  can  bee seen  in
Fig.~ref{fg4}(b),  but  the   value  deviates   significantly  from
$1/(\alpha-1)$ (dotted  and dashed  red lines). In  fact, this  is not
surprising  if considering  the results  in Fig.~\ref{fg3},  where we
found  that the  scaling of  finite-time averages  on  the logarithmic
scale   agrees  rather   well   with  the   prediction  for   infinite
trajectories, suggesting that the correction terms effectively cause a
rescaling  of  the   generalized  time-average  diffusion
coefficient. We might  call this \emph{apparent ultraweak ergodicity
breaking}.

We investigated the ultraweakly ergodic behavior of superdiffusive LWs,
finding a pronounced scatter of apparent scaling exponents of the time
averaged MSD for finite-time trajectories. These apparent scaling
exponents range between ballistic motion (sticking to one velocity mode)
down to subdiffusive values (localization due to erratic hopping between
different velocity modes). Moreover, averaged over many
individual trajectories, the time averaged MSD is pronouncedly smaller
than for very long trajectories. We quantify these effects in terms of
an ergodicity breaking parameter.

The present results reveal the importance to take into account the effects
of the finiteness of
trajectories when interpreting experimental results. They also demonstrate
how the measured time series of different lengths reveal more reliable
information  about the fundamental  underlying dynamical  process. The
additional  information comes  from  the dependence  of time  averaged
quantities  on  the  length  of  time series.  Instead  of attempting
to measure or generate longer and longer time series to  extract reliable
time averaged quantities,  one could instead use many shorter time series
and obtain even more reliable results.
Our results may also provide  an alternative and more robust method of
determining    exponents   of    probability    densities   of    step
durations. Namely,  since we inevitably  expect poor sampling  of very
long   events    this   might    be   reflected   in    the   obtained
exponent.  Using the  time averaged MSD from  measurements of different
(but known) durations one  should in  principle be  able to  determine the
exponent more accurately.

We acknowledge funding from the Academy of Finland (FiDiPro scheme) and
the German Ministry for Science and Education.


\begin{thebibliography}{00}

\bibitem{sims} D. W. Sims et al., Nature \textbf{451}, 1098 (2008);
N. E. Humphries et al., Nature \textbf{465}, 1066 (2010).

\bibitem{nathan} R. Nathan et al., Proc. Natl. Acad. Sci. USA \textbf{105},
19052 (2008).

\bibitem{bara} M. C. Gonz{\'a}lez, C. A. Hidalgo, and A.-L. Barab{\'a}si,
Nature \textbf{453}, 779 (2008); D. Brockmann, Phys. World (2), 31 (2010).

\bibitem{spider} G. Ramos-Fernandez et al., Behav. Ecol. Sociobiol. \textbf{55},
223 (2003).

\bibitem{Zumofen1} G. Zumofen and J. Klafter, Phys. Rev. E {\bf 47}, 851 (1993).

\bibitem{Zumofen2} G. Zumofen and J. Klafter, Physica D {\bf 69}, 436 (1993).

\bibitem{geisel0} T. Geisel, S. Thomae, Phys. Rev. Lett. {\bf 52},
1936 (1984).

\bibitem{swinney} T. H. Solomon, E. R. Weeks, and H. L. Swinney, Phys. Rev.
Lett. \textbf{71}, 3975 (1993); G. Zumofen and J. Klafter, Phys. Rev. E
\textbf{51}, 1818 )1995).

\bibitem{Geisel} T. Geisel, J. Nierwetberg, and A. Zacherl,
Phys. Rev. Lett. {\bf 54}, 6161 (1985).

\bibitem{Geisel2} R. Fleischmann, T. Geisel, and R. Ketzmerick,
Europhys. Lett. {\bf 25}, 219 (1994).

\bibitem{Marksteiner} S. Marksteiner, K. Ellinger, and P. Zoller,
 Phys. Rev. A {\bf 53}, 3409 (1996). 

\bibitem{wiersma} P. Barthelemy, J. Bertolotti, and D. S. Wiersma, Nature
\textbf{453}, 495 (2008).

\bibitem{Silbey} Y.-J. Jung, E. Barkai, and R. J. Silbey, Chem. Phys. {\bf
284}, 181 (2002).

\bibitem{deJager} M. de Jager, F. J. Weissing, P. M. J. Herman,
B. A. Nolet, and J. van de Koppel, Science {332}, 1551 (2011).

\bibitem{Harris} T. H. Harris {\it et al.}, Nature {\bf 486}, 545 (2012).

\bibitem{pt} E. Barkai, Y. Garini, and R. Metzler, Phys. Today \textbf{65}(8),
29 (2012).

\bibitem{yonghe} Y. He, S. Burov, R. Metzler, and E. Barkai, Phys. Rev. Lett.
{\bf 101}, 058101 (2008); S. Burov, J.-H. Jeon, R. Metzler, and E. Barkai,
Phys. Chem. Chem. Phys. \textbf{13}, 1800 (2011).

\bibitem{deng} W. Deng and E. Barkai, Phys. Rev. E \textbf{79}, 011112 (2009);
I. Goychuk, \emph{ibid.} \textbf{80}, 046125 (2009); J.-H. Jeon and R. Metzler,
\emph{ibid.} \textbf{81}, 021103 (2010).

\bibitem{jae} Note that even ergodic anomalous diffusion processes may show
non-ergodic features under confinement, see J.-H. Jeon, R. Metzler, Phys.
Rev. E {\bf 85}, 021147 (2012).

\bibitem{weigel} A. V. Weigel \emph{et al.}, Proc. Nat. Acad. Sci. USA
\textbf{108}, 6438 (2011).

\bibitem{lene} J.-H. Jeon \emph{et al.}, Phys. Rev. Lett. \textbf{106}, 048103
(2011).

\bibitem{shleklawong} M. F. Shlesinger, J. Klafter, and Y. M. Wong, J. Stat.
Phys. \textbf{27}, 499 (1982).

\bibitem{montroll} E. W. Montroll and G. H. Weiss, J. Math. Phys. \textbf{10},
753 (1969); H. Scher and E. W. Montroll, Phys. Rev. B \textbf{12}, 2455 (1975).

\bibitem{klablushle} J. Klafter, A. Blumen, and M. F. Shlesinger, Phys. Rev. A
{\bf 35}, 3081 (1987).

\bibitem{Klafter2} J. Klafter, M. F. Shlesinger, and G. Zumofen,
Phys. Today {\bf 49}, 33 (1996).

\bibitem{report} R. Metzler and J. Klafter, Phys. Rep. \textbf{339}, 1
(2000); J. Phys. A \textbf{37}, R161 (2004).

\bibitem{Trefan} G. Tref\'an, E. Floriani, B. J. West, and P. Grigolini,
 Phys. Rev. E {\bf 50}, 2564 (1994).

\bibitem{Ralf15} I. M. Sokolov and R. Metzler, Phys. Rev. E {\bf 67},
 010101 (2003).

\bibitem{Marcin} M. Magdziarz, W. Szczotka, and Zebrowski, J. Stat. Phys.
{\bf 147}, 74 (2012). 

\bibitem{Green_Kubo} R. Kubo, Rep. Prog. Phys. {\bf 29}, 255 (1966);
J.-P. Hansen, I.R. McDonald, {\it Theory of simple liquids,
3rd. Ed.}, (Academic Press, Amsterdam, 2006).

\bibitem{REM} Note the difference to the definition of the ergodicity breaking
parameter $\mathrm{EB}$ introduced previously \cite{yonghe,pt}.

\end{thebibliography}
\end{document}